\documentclass[useAMS,usenatbib,usegraphicx]{mn2e}

\title[X-rays and Young SNe expansion]{What are Published X-ray
  lightcurves telling us about Young Supernova Expansion?}
\author[Dwarkadas \& Gruszko]{V. V. Dwarkadas,$^{1}$\thanks{E-mail:
    vikram@oddjob.uchicago.edu} and
  J. Gruszko,$^{2}$\\ $^{1}$Department of Astronomy and Astrophysics,
  U Chicago, 5640 S Ellis Ave, Chicago, IL 60637\\ $^{2}$Department of
  Physics and Astronomy, University of Rochester, Rochester, NY
  14627-0171}
  
\begin{document}
\newcommand{\vper}{\mbox{${v_{\perp}}$}}
\newcommand{\vpar}{\mbox{${v_{\parallel}}$}}
\newcommand{\uper}{\mbox{${u_{\perp}}$}}
\newcommand{\vperout}{\mbox{${{v_{\perp}}_{o}}$}}
\newcommand{\uperout}{\mbox{${{u_{\perp}}_{o}}$}}
\newcommand{\vperin}{\mbox{${{v_{\perp}}_{i}}$}}
\newcommand{\uperin}{\mbox{${{u_{\perp}}_{i}}$}}
\newcommand{\upar}{\mbox{${u_{\parallel}}$}}
\newcommand{\uparout}{\mbox{${{u_{\parallel}}_{o}}$}}
\newcommand{\vparout}{\mbox{${{v_{\parallel}}_{o}}$}}
\newcommand{\uparin}{\mbox{${{u_{\parallel}}_{i}}$}}
\newcommand{\vparin}{\mbox{${{v_{\parallel}}_{i}}$}}
\newcommand{\dout}{\mbox{${\rho}_{o}$}}
\newcommand{\din}{\mbox{${\rho}_{i}$}}
\newcommand{\da}{\mbox{${\rho}_{1}$}}
\newcommand{\mfast}{\mbox{$\dot{M}_{f}$}}
\newcommand{\mslow}{\mbox{$\dot{M}_{a}$}}
\newcommand{\beqn}{\begin{eqnarray}}
\newcommand{\eeqn}{\end{eqnarray}}
\newcommand{\be}{\begin{equation}}
\newcommand{\ee}{\end{equation}}
\newcommand{\noi}{\noindent}
\newcommand{\ftheta}{\mbox{$f(\theta)$}}
\newcommand{\gtheta}{\mbox{$g(\theta)$}}
\newcommand{\ltheta}{\mbox{$L(\theta)$}}
\newcommand{\stheta}{\mbox{$S(\theta)$}}
\newcommand{\utheta}{\mbox{$U(\theta)$}}
\newcommand{\xitheta}{\mbox{$\xi(\theta)$}}
\newcommand{\vs}{\mbox{${v_{s}}$}}
\newcommand{\ro}{\mbox{${R_{0}}$}}
\newcommand{\pa}{\mbox{${P_{1}}$}}
\newcommand{\va}{\mbox{${v_{a}}$}}
\newcommand{\vo}{\mbox{${v_{o}}$}}
\newcommand{\vp}{\mbox{${v_{p}}$}}
\newcommand{\vw}{\mbox{${v_{w}}$}}
\newcommand{\vf}{\mbox{${v_{f}}$}}
\newcommand{\lprime}{\mbox{${L^{\prime}}$}}
\newcommand{\uprime}{\mbox{${U^{\prime}}$}}
\newcommand{\sprime}{\mbox{${S^{\prime}}$}}
\newcommand{\xiprime}{\mbox{${{\xi}^{\prime}}$}}
\newcommand{\mdot}{\mbox{$\dot{M}$}}
\newcommand{\msun}{\mbox{$M_{\odot}$}}
\newcommand{\yr}{\mbox{${\rm yr}^{-1}$}}
\newcommand{\kms}{\mbox{${\rm km} \;{\rm s}^{-1}$}}
\newcommand{\lambdav}{\mbox{${\lambda}_{v}$}}
\newcommand{\lequ}{\mbox{${L_{eq}}$}}
\newcommand{\eqpratio}{\mbox{${R_{eq}/R_{p}}$}}
\newcommand{\ra}{\mbox{${r_{o}}$}}
\newcommand{\bfig}{\begin{figure}[h]}
\newcommand{\efig}{\end{figure}}
\newcommand{\tone}{\mbox{${t_{1}}$}}
\newcommand{\done}{\mbox{${{\rho}_{1}}$}}
\newcommand{\dsn}{\mbox{${\rho}_{SN}$}}
\newcommand{\dzero}{\mbox{${\rho}_{0}$}}
\newcommand{\ve}{\mbox{${v}_{e}$}}
\newcommand{\vej}{\mbox{${v}_{ej}$}}
\newcommand{\Mch}{\mbox{${M}_{ch}$}}
\newcommand{\mej}{\mbox{${M}_{e}$}}
\newcommand{\Mst}{\mbox{${M}_{ST}$}}
\newcommand{\dam}{\mbox{${\rho}_{am}$}}
\newcommand{\Rst}{\mbox{${R}_{ST}$}}
\newcommand{\Vst}{\mbox{${V}_{ST}$}}
\newcommand{\Tst}{\mbox{${T}_{ST}$}}
\newcommand{\no}{\mbox{${n}_{0}$}}
\newcommand{\Efif}{\mbox{${E}_{51}$}}
\newcommand{\rsh}{\mbox{${R}_{sh}$}}
\newcommand{\msh}{\mbox{${M}_{sh}$}}
\newcommand{\vsh}{\mbox{${V}_{sh}$}}
\newcommand{\vrev}{\mbox{${v}_{rev}$}}
\newcommand{\rpr}{\mbox{${R}^{\prime}$}}
\newcommand{\mpr}{\mbox{${M}^{\prime}$}}
\newcommand{\vpr}{\mbox{${V}^{\prime}$}}
\newcommand{\tpr}{\mbox{${t}^{\prime}$}}
\newcommand{\cone}{\mbox{${c}_{1}$}}
\newcommand{\ctwo}{\mbox{${c}_{2}$}}
\newcommand{\cthree}{\mbox{${c}_{3}$}}
\newcommand{\cfour}{\mbox{${c}_{4}$}}
\newcommand{\Te}{\mbox{${T}_{e}$}}
\newcommand{\Ti}{\mbox{${T}_{i}$}}
\newcommand{\Ha}{\mbox{${H}_{\alpha}$}}
\newcommand{\Rprime}{\mbox{${R}^{\prime}$}}
\newcommand{\Vprime}{\mbox{${V}^{\prime}$}}
\newcommand{\Tprime}{\mbox{${T}^{\prime}$}}
\newcommand{\Mprime}{\mbox{${M}^{\prime}$}}
\newcommand{\rprime}{\mbox{${r}^{\prime}$}}
\newcommand{\rfprime}{\mbox{${r}_f^{\prime}$}}
\newcommand{\vprime}{\mbox{${v}^{\prime}$}}
\newcommand{\tprime}{\mbox{${t}^{\prime}$}}
\newcommand{\mprime}{\mbox{${m}^{\prime}$}}
\newcommand{\Me}{\mbox{${M}_{e}$}}
\newcommand{\nh}{\mbox{${n}_{H}$}}
\newcommand{\rr}{\mbox{${R}_{2}$}}
\newcommand{\rf}{\mbox{${R}_{1}$}}
\newcommand{\vtwo}{\mbox{${V}_{2}$}}
\newcommand{\vout}{\mbox{${V}_{1}$}}
\newcommand{\dshell}{\mbox{${{\rho}_{sh}}$}}
\newcommand{\dwind}{\mbox{${{\rho}_{w}}$}}
\newcommand{\dslow}{\mbox{${{\rho}_{s}}$}}
\newcommand{\dfast}{\mbox{${{\rho}_{f}}$}}
\newcommand{\vfast}{\mbox{${v}_{f}$}}
\newcommand{\vslow}{\mbox{${v}_{s}$}}
\newcommand{\cc}{\mbox{${\rm cm}^{-3}$}}
\newcommand{\apj}{\mbox{ApJ}}
\newcommand{\apjl}{\mbox{ApJL}}
\newcommand{\apjs}{\mbox{ApJS}}
\newcommand{\aj}{\mbox{AJ}}
\newcommand{\araa}{\mbox{ARAA}}
\newcommand{\nat}{\mbox{Nature}}
\newcommand{\aap}{\mbox{AA}}
\newcommand{\gca}{\mbox{GeCoA}}
\newcommand{\pasp}{\mbox{PASP}}
\newcommand{\mnras}{\mbox{MNRAS}}
\newcommand{\apss}{\mbox{ApSS}}

\date{\today}

\pagerange{\pageref{firstpage}--\pageref{lastpage}} \pubyear{2010}

\maketitle

\label{firstpage}

\begin{abstract}
Massive stars lose mass in the form of stellar winds and
outbursts. This material accumulates around the star. When the star
explodes as a supernova (SN) the resulting shock wave expands within
this circumstellar medium. The X-ray emission resulting from the
interaction depends, {among other parameters}, on the density of
this medium, and therefore the variation in the X-ray luminosity can
be used to study the variation in the density structure of the
medium. In this paper we explore the X-ray emission and lightcurves of
all known SNe, in order to study the nature of the medium into which
they are expanding. In particular we wish to investigate whether young
SNe are expanding into a steady wind medium, as is most often assumed
in the literature. We find that in the context of the theoretical
arguments that have been generally used in the literature, many young
SNe, and especially those of Type IIn, which are the brightest X-ray
luminosity class, do not appear to be expanding into steady
winds. Some IIns appear to have very steep X-ray luminosity declines,
indicating density declines much steeper than r$^{-2}$. However, other
IIns show a constant or even increasing X-ray luminosity {over
  periods of months to years}. Many other SNe do not appear to have
declines consistent with expansion in a steady wind. SNe with lower
X-ray luminosities appear to be more consistent with steady wind
expansion, although the numbers are not large enough to make firm
statistical comments. The numbers do indicate that the expansion and
density structure of the circumstellar medium must be investigated
before assumptions can be made of steady wind expansion. Unless a
steady wind can be shown, mass-loss rates deduced using this
assumption may need to be revised.

\end{abstract}


\section{Introduction}
\label{sec:intro}

In 1972, in a review article on supernova remnants (SNRs),
\citet{woltjer72} did not devote much attention to the ambient medium
into which the SN was expanding, suggesting that the ``importance of
the swept-up matter is comparatively minor'' in the ejecta-dominated
stage, as ``everything depends on the details of the explosive
process''. \citet{chevalier77} also considered the interaction of SNRs
only with the interstellar medium. However, by 1982 the idea appears
to have been established that core-collapse supernovae (SNe) expand
not into the interstellar medium but the wind-blown circumstellar
medium (CSM) ejected by the progenitor star. \citet{chevalier1982a,
  chevalier1982b} derived self-similar solutions for this
interaction. Since then this concept has been widely tested and
accepted, so much so that nowadays it is commonly accepted that young
core-collapse SNe expand into the wind-blown structures created by
their progenitor stars, and the resulting X-ray emission is due to the
forward and reverse shocks interacting with the ambient medium
\citep{chevalier1994}.

The wind from a star is defined basically by two parameters, the
mass-loss rate and the wind velocity. If these parameters are constant
it is referred to as a steady wind.  Another common paradigm that
frequently now appears in the literature is that young core-collapse
SNe are expanding into steady winds. This has been used to compute the
mass-loss rates of X-ray supernovae \citep{Immler2003, ik05}, and
especially optical SNe \citep{smithetal08, scsff10, kieweetal10}.  In
very few cases has an attempt been made to determine whether the wind
is in fact steady, and some of the calculations were shown to be
inconsistent in this respect with the results obtained
\citep{dwarkadas11}. However, a steady wind is a reasonable
approximation that makes analytical calculations possible, whereas the
alternative is to do much more complicated calculations which may not
even be possible without substantially more data.

A steady wind with a constant mass-loss rate ($\dot{M}$) and wind
velocity ($v_w$) results in a density $\rho = \mdot/(4 \pi r^2
v_w)$. The density profile thus evolves as r$^{-2}$. If the wind
parameters vary with time, the density profile would deviate from an
r$^{-2}$ evolution. Thus by determining the nature of the density
profile into which young core-collapse SNe are expanding, one could
infer the variation in wind parameters (if any), and the nature of the
progenitor mass-loss. Supernova expansion into the surrounding medium
allows a natural probe into the density structure of this
medium. Since the SN expansion velocity can be about 10-1000 times the
velocity of the surrounding medium, depending on whether the
progenitor was a Wolf-Rayet (W-R) star or a red supergiant (RSG) star,
the SN ejecta allows us to probe 10-1000 years of wind evolution in 1
year of SN evolution. 

The X-ray emission resulting from the circumstellar interaction is
proportional to the density structure (see \S \ref{sec:xraysne}), and
therefore the evolution of the X-ray light curve can be used to probe
the variations in density structure in the ambient
medium. \citet{chevalier1982b} studied the X-ray emission from
core-collapse SNe, and formulated an expression for their X-ray
luminosity in terms of the evolutionary parameters. This has since
been used to describe the X-ray light curves of
SNe. \citet{schlegel95} reviewed X-ray observations of all 5 X-ray SNe
known till that time. The pace of discovery has picked up with the
advent of {\it Chandra}, {\it XMM-Newton} and {\it Swift}, with
\citet{Immler2003} listing 15 SNe, \citet{schlegel06} having 25 on his
list, and \citet{immler07} listing 32 SNe. \footnote{ An extremely
  useful compilation of X-ray SNe, maintained by Dr.~Stefan Immler, is
  available at the following web site:
  lheawww.gsfc.nasa.gov/users/immler/supernovae\_list.html}

In this paper we aim to use published X-ray light-curves of SNe to
explore the nature of the medium into which they are evolving. We will
explore what can be learned from X-ray light curves in \S
\ref{sec:xraysne}, and the limitations of this technique. In the next
section \ref{sec:xraylc} we show the lightcurves of X-ray SNe and fits
to the data, and discuss the resulting implications. We analyze these
results for individual SNe and SNe types in \S
\ref{sec:analysis}. Finally, in \S \ref{sec:summary} we summarize our
results and discuss their consequences.

\section[]{What can we learn from X-ray light-curves of SNe?}
\label{sec:xraysne}

A very thorough discussion of the evolution of the X-ray light curve
from young SNe was carried out by \citet{flc96}, in reference to SN
1993J. In this discussion, and throughout this paper, we will make use
of many of the results described therein. For the sake of
completeness, we present here a simplified analysis with basic results
that develops and summarizes the most important ideas. This summary
follows the description given in \citet{ddb10}.

The expansion of a SN shock wave into the ambient medium gives rise to
a forward and reverse-shocked structure, separated by a contact
discontinuity \citep{chevalier1982a}. The shocks heat up the medium to
high temperatures, producing X-ray emission.  The X-ray luminosity
$L_x$ from a source depends on the electron density $n_e$, the
emitting volume $V$ and the cooling function $\Lambda$

\be
 L_x \sim {n_e}^2 \, \Lambda \, V
\ee

The density of the medium into which the SN shock wave is expanding is
assumed to go as $r^{-s}$, with s=2 for a steady wind. The emission
arises from a thin shell of radius $\Delta r$ in between the forward
and reverse shocks, at a mean radius $r$ , whose volume $V$ can be
expressed as $4 \pi r^2\, \Delta r$. Although the wind may not have
been steady, the evolution will be self-similar as long as both the SN
ejecta and the circumstellar medium density profiles are described by
power-laws \citep{chevalier1982a}. In the self-similar case, $\Delta r
\propto r$, and therefore $V \propto r^3$. For a SN in the early
stages, the post-shock temperature is going to be much larger than
10$^7$K, and the cooling function is assumed to vary as $T^{0.5}$
\citep{chevalier1994}. For a strong shock, $T \propto {v_s}^2$, and
therefore the cooling function $\Lambda \propto v_s \propto r/t$ in
the self-similar case.

Therefore we get 

\be 
L_x \sim\; r^{-2s} \,\frac{r}{t} \, r^3
\ee

which gives

\be 
\label{eq:lum}
L_x \sim \; \frac{r^{4 - 2s}}{t}
\ee

For s=2 this gives the well-known result that the emission decreases
inversely with time, $L_x \propto t^{-1}$. This result is inherent in
the formula for mass-loss derived by \citet{Immler2003}.

In the self-similar case, with $\rho_{ej} \propto A\; t^{-3}\; v^{-n}$
and $\rho_{cs} \propto r^{-s}$, the radius of the SN will evolve as
\citep{chevalier1982a} 

\be
\label{eq:radexp}
R_{SN} \propto t^{\alpha} ~~~~\alpha=(n-3)/(n-s)
\ee

\noi 
Using this in equation \ref{eq:lum} gives us that in a
general case,

\be 
\label{eq:lumgen}
L_x \sim \; t^{-(12-7s+2ns-3n)/(n-s)}
\ee

This is a general formula that has often been used to describe the
evolution of the X-ray luminosity. An important caveat must be pointed
out here - this formula considers the total X-ray emission over all
temperatures. However X-ray satellites measure the emission only in a
very narrow band. {\it ROSAT} covered the region between 0.4 - 2.4
keV. {\it Chandra} and {\it XMM-Newton} cover a slightly larger band,
around 0.4-10/15 keV, but the effective area is much higher at lower
energies. Therefore, although equation \ref{eq:lumgen} shows how the
X-ray luminosity will vary with time over the {\em entire} X-ray
range, it is doubtful that this variation could be observed. We
observe the variation only for that part of the luminosity that falls
within a given X-ray mission's observable band.  This would have the
same time dependence if and only if the ratio of the luminosity in
that discrete band to that of the total X-ray region is constant.

However this is not likely. As the SN expands, we expect that the
shock will sweep up more material and decelerate. The post-shock
temperature, proportional to the square of the shock velocity, will
decrease with time, and the X-ray emission would gradually shift from
high to lower temperatures. Since X-ray satellites can detect emission
mainly in the softer bands, we would expect that the luminosity in the
lower energy range would increase at the expense of that in the higher
energy range. Thus, the ratio of X-ray emission we observe to the
total should increase over time, leading to a time dependence flatter
than the overall time dependence in a steady wind of $t^{-1}$.

In some cases, this can be derived. From above, for a non-radiative
shock, the shock velocity goes as $v_{sh} \propto t^{\alpha -1}$, and
the shock temperature therefore decreases as $T_{sh} \propto
t^{2(\alpha-1)}$. As shown in \citet{flc96}, the luminosity in a given
energy band with E $<< kT_{sh}$ will then go as:

\be 
\label{eq:lumband}
L_x(E) \sim \; t^{-(6-5s+2ns-3n)/(n-s)} \sim \; t^{-\beta}
\ee

Compared to equation \ref{eq:lumgen}, note that this gives a flatter
time dependence, as expected. For $s=2$ we get that $L_x \sim
t^{-(n-4)/(n-2)}$. The index $\beta$ is now always $< 1$, implying
that the time dependence we see will not be $t^{-1}$ but will be
flatter, and is now a function of the ejecta density power-law. For $n
> 5$, as required by the Chevalier solution, $0.33 < \beta < 1$. For
commonly used values n=9, $\beta=0.714$, and for n=11,
$\beta=0.77$. Therefore, even for a steady wind, we are more likely to
see $L_x \sim t^{-(0.75 \pm 0.05)}$ . It should be noted that for
$s=2$, we will not see a steeper luminosity dependence than $t^{-1}$,
or flatter than $t^{-0.33}$.

For $s \neq 2$ the situation is more complex. For $s=1$, we get
$\beta=-1$, i.e.~the flux is linearly increasing with time,
irrespective of the value of $n$. Thus a decreasing wind density can
still give rise to an increasing flux. For $s=1.5$ we get that
$\beta=-1.5/(n-1.5)$ is always negative, and therefore that the
luminosity is increasing with time, whereas the total X-ray luminosity
from equation \ref{eq:lumgen} is actually decreasing with time. Thus
this clearly shows the difference between what we would observe with
X-ray satellites, and what the total X-ray luminosity is doing.

Since we can get both decreasing and increasing luminosity dependence
with time, it follows that for some specific values we could observe a
constant luminosity with time, i.e.~$\beta=0$. This depends on the
value of the ejecta density. For $n=9$ we obtain from equation
\ref{eq:lumband} that $\beta=(13s-21)/(9-s)$. Therefore, for $s=21/13
\sim 1.62$ the luminosity would be constant with time.

If the temperature of one of the shocks, especially the reverse shock,
is lower than about 3 $\times 10^7$, then the cooling function
behavior changes to $\Lambda \propto T^{-0.6}$
\citep{chevalier1994}. For simplicity and to illustrate the difference
in evolution, we may assume that the dependence goes as
$T^{-0.5}$. Then we get, following the argument above, that the
$\Lambda$ dependence can be inverted, giving

\be 
L_x \sim\; r^{-2s} \,\frac{t}{r} \, r^3
\ee

which leads to 

\be 
\label{eq:lum}
L_x \sim \; {r^{2 - 2s}t}
\ee

and therefore, assuming that $r \propto t^{\alpha}$, we get

\be 
\label{eq:lumbandlow}
L_x(E) \sim \; t^{-(2ns-5s-3n+6)/(n-s)} \sim \; t^{-\beta}
\ee

The situation becomes more complicated if the shock becomes
radiative. This has also been discussed by \citet{flc96}. In this case
the luminosity goes as

\be
\label{eq:lumcool}
L_x \sim \; t^{-(15-6s+ns-2n)/(n-s)}
\ee

while, as long as the cool shell is optically thick, the luminosity in
a band evolves as

\be
\label{eq:lumcoolband}
L_x(E) \sim \; T_e^{0.16}t^{(3-2s)(n-3)/(n-s)}
\ee

where the luminosity is now proportional also to the electron
temperature $T_e$ with time. It is interesting to note here that for a
density gradient $s=1.5$ the flux is constant irrespective of the
value of the ejecta gradient $n$. For $s > 1.5$ the luminosity is
decreasing, while for $s < 1.5$ it is increasing. Note that for a
steady wind with $s=2$ the time dependence $\beta = \alpha$, i.e. the
luminosity evolution has the same time dependence as the radius
evolution without cooling, although with the opposite sign.

We emphasize that the light curves of young SNe expanding into a
steady wind, {\em observed in a narrow band}, would not be expected to
have the canonical $t^{-1}$ dependence that so many have postulated
\citep{pooleyetal02, immleretal02, Immler2003, ik05, sp08}, but would
have a flatter dependence that is a function of the ejecta
power-law. This has been very nicely shown by \citet{flc96} but is
routinely ignored.

Even though this approximate theory works nicely in some cases, it is
important that we remain cognizant of its limitations. This includes
the fact that a power-law profile is assumed for both the ejecta and
the surrounding wind medium; that self-similarity is assumed; and that
the same cooling function is implied for both forward and reverse
shocks. It is possible that one or both of the ejecta and CSM density
profiles do not decrease as a power-law; that the solution does not
depict self-similar behaviour, and that different temperatures result
in different cooling functions at each shock. While in the later case
the result for each shock could possibly be worked out in a similar
manner, the total X-ray emission from the SN would depend on details
such as how much was emanating from each shock in which temperature
range, and would not be easy to compute analytically. Finally, such a
theory computes the emission due only to circumstellar interaction,
and does not consider X-ray emission due to other mechanisms such as
Inverse Compton scattering or a central pulsar, or due to other
components such as radiative shocks in a clumpy medium. Thus, although
there is a wide range over these ideas are applicable, it is prudent
to keep the shortcomings in mind.

\section{Observed X-ray light curves of young core-collapse SNe}
\label{sec:xraylc}

\subsection{X-ray Data:}

In order to study X-ray SNe light curves, we have compiled data on SNe
that was available in the literature, using published values of
luminosity/flux. The time after outburst is also taken from the
literature, although in cases where different times of explosion were
listed for the same SNe in different papers, we have appealed to the
optical light curve for clarification. Table 1 lists the various
papers from which the data were compiled. In some cases the data were
not explicitly listed in a table but had to be read off a figure,
which introduced further error in the numbers. In a few cases (SN
2005kd, SN 2006jd), we have downloaded the publicly archived data-set
and computed the flux, in order to procure at least one more
data-point so that we may construct a light curve. In one case (SN
1993J) the most current data point is from a {\it Chandra} observation
on which the first author is PI. We have chosen not to include
GRB-related SNe in this compilation, as their X-ray emission is not
expected to be due purely to circumstellar interaction. We have also
not included SN 2008D, a very well observed SN whose early lightcurve
at least was attributed to shock breakout \citep{soderbergetal08} and
thus did not fit in with the circumstellar model discussed herein.

\begin{table*}
\begin{minipage}{126mm}
\caption{The list of core-collapse supernovae for which X-ray data
  were available and used in this paper. The source of the data is
  listed, along with the energy range for the luminosity. DG11 refers
  to this paper. } \centering
  \begin{tabular}{l c r}
  \hline
   SN Name  & Energy Range (kev)& Publication\\
\hline
1986J & .5-2.5 & \citet{houck05} \\
1988Z & .2-2.0 & \citet{sp06} \\
1995N & .1-2.4 & \citet{chandraetal05} \\
1998S & .4-2.0 & \citet{pooleyetal02} \\
1978K & .5-2.0 & \citet{schlegel04, ls07} \\
1993J & .5-2.4 & \citet{chandraetal09}, DG11  \\
1979C & .3-2.0 & \citet{patnaudeetal11} \\
2002ap & .3-2.0 & \citet{sutariaetal03} \\
1994W & .1-2.4 & \citet{schlegel99} \\
1996cr & .5-2.0 & \citet{baueretal2008} \\
1981K & .5-2.0 & \citet{ilyw07} \\
1987A & .5-2.0 & \citet{Hasinger1996, parketal07} \\
2010F & .2-10 & \citet{russelletal10} \\
2009mk & .2-10 & \citet{ri10} \\
2009gj & .2-10 & \citet{ir09} \\
2009dd & .2-10 & \citet{irb09} \\
2008ij & .2-10 & \citet{ipbm09} \\
2008ax & .2-10 & \citet{immler08} \\
2008M & .2-10 & \citet{immler10a} \\
2006jd & .2-10 & \citet{ibfp07}, DG11 \\
2006jc & .2-10 & \citet{immleretal08} \\
2005kd & .2-10 & \citet{ipb07, pif07}, DG11 \\
2005ip & .2-10 & \citet{ip07} \\
2004et & .4-8 & \citet{rjcc07} \\
2004dj & .5-8 & \citet{pl04} \\
2002hi & .5-10 & \citet{pl03} \\
2002hh & .4-8 & \citet{pl02} \\
2001ig & .2-10 & \citet{sr02} \\
2001gd & .3-5 & \citet{perezetal05} \\
1999em & .5-8 & \citet{pooleyetal02} \\
2003L & .5-5 & \citet{soderbergetal05} \\
2003bg & .3-10 & \citet{soderbergetal06} \\
1999gi & .3-10 & \citet{schlegel01} \\
1970G & .3-2 & \citet{ik05} \\
2006bp & .2-10 & \citet{ib06, immleretal07} \\
1941C & .3-8 & \citet{sp08} \\
1959D & .3-8 & \citet{sp08} \\
1968D & .3-8 & \citet{sp08} \\
1980K & .5-4 & \citet{schlegel95, sp08} \\
2010jl & 0.2-10 & \citet{imp10} \\
2010jr & 0.2-10 & \citet{irrp10} \\
\hline
\end{tabular}
\end{minipage}
\end{table*}

\begin{figure*}
\includegraphics[angle=90,scale=0.5]{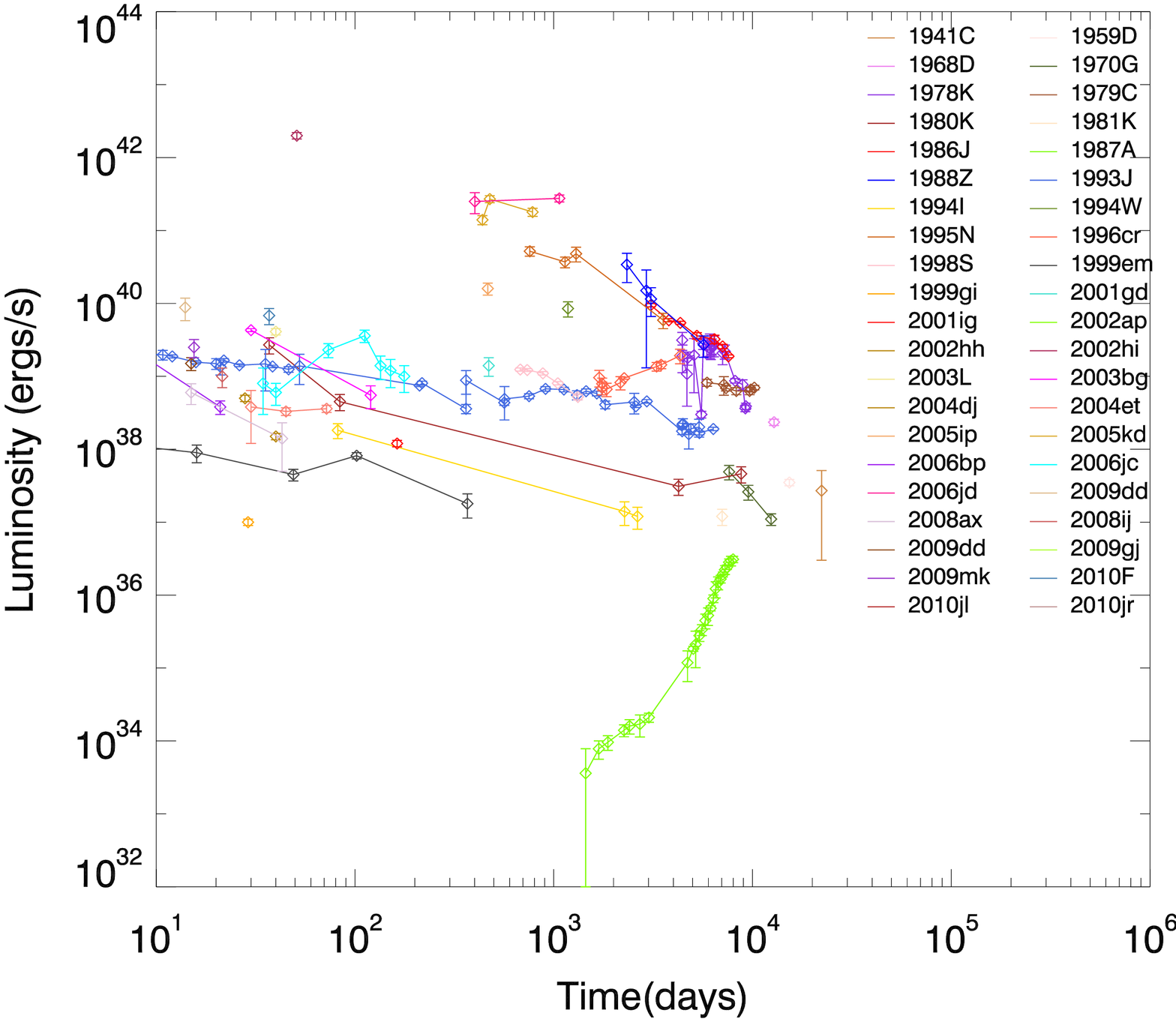}
\caption{The X-ray lightcurves of all observed X-ray SNe found in the
  literature, with one or more data points. The observed range of
  luminosities spans about 5 orders of magnitude, from 10$^{37}$ to
  10$^{42}$ ergs s$^{-1}$, excluding SN 1987A, which is a very low
  luminosity SN that was observed only because of its proximity.
\label{fig:plotlxall}}
\end{figure*}

Figure \ref{fig:plotlxall} plots the datapoints for all X-ray observed
SNe that we found in the literature with at least 1 published data
point. The number is only 42, although it is rapidly increasing,
thanks mainly to {\it Swift}. We plot light curves for almost all
published SNe with multiple data points, enabling an X-ray light curve
to be constructed. As far as possible we have constructed the light
curve in the 0.4-2.4 keV or related energy range, which usually has
the most data for SNe that exploded during or before the {\it ROSAT}
era, and thus ensures the largest duration light curves. In some
cases, as mentioned in Table 1, a larger energy band is used. Most of
these are SNe observed with {\it Swift}. All points on a lightcurve
lie in the same energy range, which is given in Table 1. In order to
accomplish this using only published data, we had to select the energy
range in which the maximum number of SN fluxes were quoted, which
sometimes meant not using available flux results lying outside the
energy range. Given our intent of assessing what the published light
curves are telling us, no attempt was made to interpolate or
extrapolate published data from a given energy range to another, which
would have required assuming a spectral model that was generally not
available.

While the number of SNe is gradually increasing, it is still not very
large, with only 20 SNe found in the literature which have flux values
at more than one epoch (although large amounts of unpublished data
exists in the archives). Given the current availability of {\it
  Chandra}, {\it XMM-Newton} and especially {\it Swift}, we hope that
this situation can be rectified in future. Only one SN (1993J) has
been observed regularly since its birth, and has the most complete
light curve, although gaps still exist. SN 1987A is probably the most
observed X-ray SN, due to its closeness and unique nature. Hard X-ray
data exist in the first 1000 days from Ginga \citep{inoueetal91}, with
progressively better resolution data thereafter from all the major
X-ray observatories, and almost every X-ray telescope, including
high-resolution gratings.

The plot shows that very few SNe possess a monotonic light curve in
the given X-ray band. This may be partially due to the published error
bars being too optimistic, but in some cases it is evident that the
light curve tends to jump around.

\subsection{Fits to the Light Curves:}

\begin{figure*}
\includegraphics[angle=90,scale=0.5]{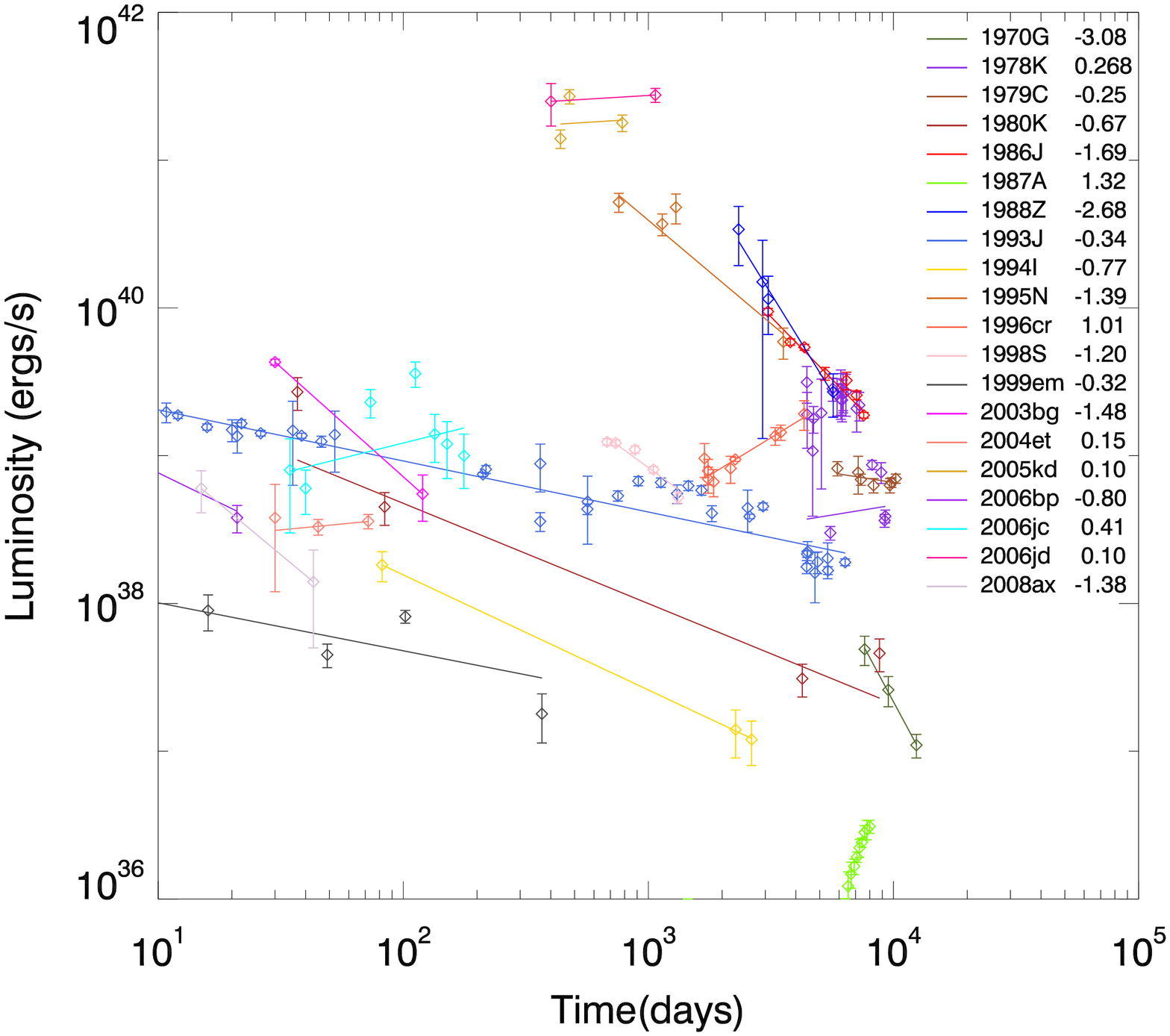}
\caption{Fits to the lightcurves of all X-ray SNe with multiple data
  points. The slope of the Luminosity v/s time curves, plotted on a
  log-log scale, is given next to each SN. The fits were calculated
  using the MPFIT routine in IDL (see text).
\label{fig:plotlxfit}}
\end{figure*}

Following the discussion in \S \ref{sec:xraysne}, we fitted the light
curves of all SNe with multiple data points (Figure
\ref{fig:plotlxall}) using an expression of the form $L_x = P(0)
\times t^{P(1)}$. The fitting was done using the MPFIT routine
\citep{markwardt09} in the IDL programming language. In Figure
\ref{fig:plotlxfit} we show the fits for various SNe. The legend gives
the value of the parameter $P(1)$ in the expression above, which
corresponds to $-\beta$ from \S \ref{sec:xraysne}.

In many cases the use of a simple power-law fitting function does not
seem appropriate, as is evident not only from the large values of
$\chi$-squared that were found, but merely by comparing the fit to the
data points. It appears that a more complicated function is needed, or
perhaps a break is needed in the power-law. In one case, SN 1978K, the
fit seems entirely off, and it is clear that something more
sophisticated is needed.

\subsection{X-ray lightcurves by SN Type:}

\begin{figure*}
\includegraphics[angle=90,scale=0.5]{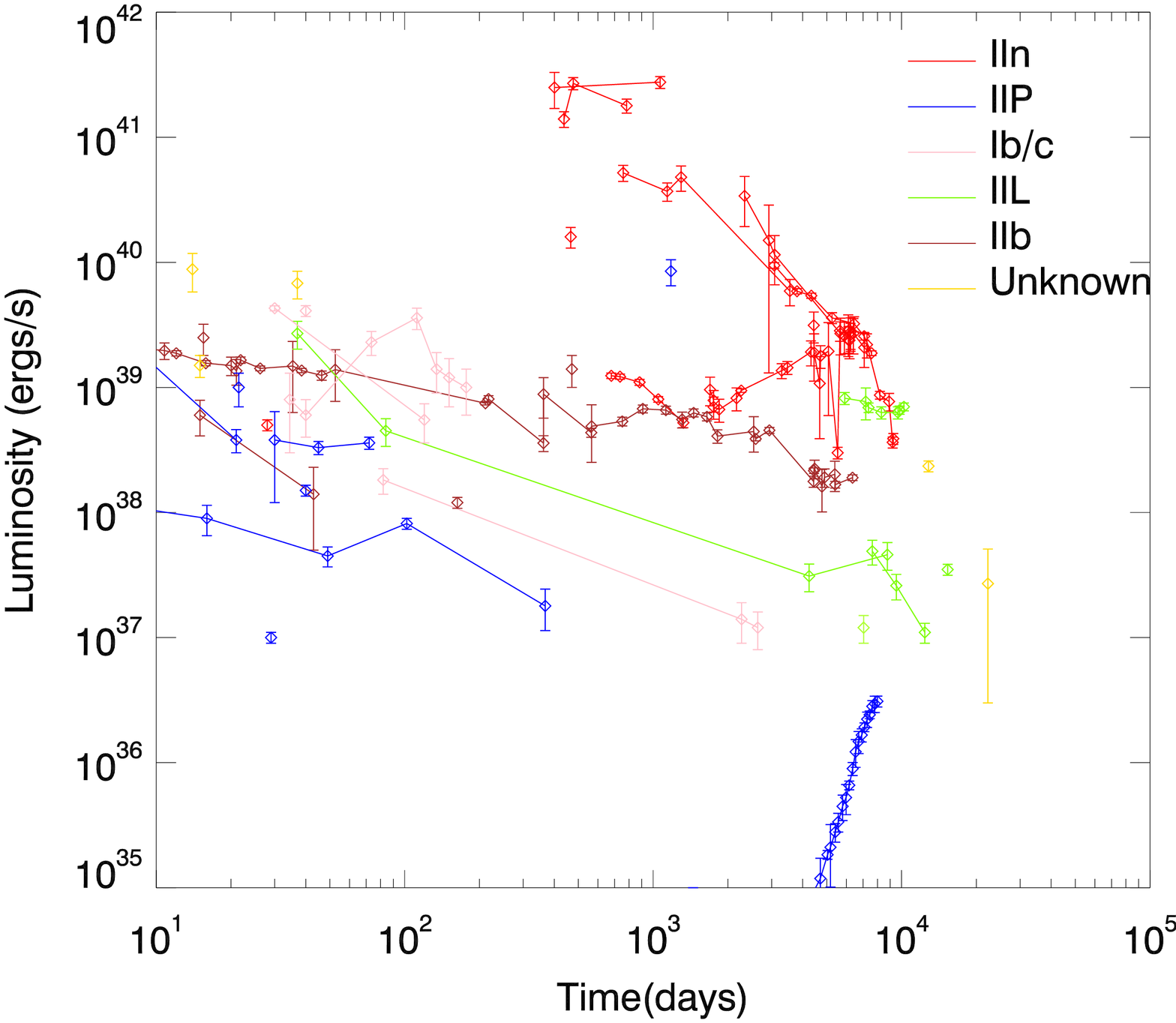}
\caption{The light curves of X-ray SNe plotted as a function of the
  type of SN. Type IIn SNe are the brightest as a group, therefore
  suggesting that they interact with regions of higher density. Others
  show much more diversity, with IIP's generally having the lowest
  luminosity as a class, although exceptions exist in all classes.
\label{fig:plotlxtype}}
\end{figure*}

In order to study the variation of the X-ray emission with type of
SNe, we show in figure \ref{fig:plotlxtype} the light curves of all
SNe plotted as a function of the type of SN. It is clear from this
plot that Type IIn SNe show the highest luminosities, and Type IIP SNe
the lowest. These bracket the Type IIL's, Type IIb's and Type Ib/c. We
caution that we have not included GRB-SNe here, and while the sample
is probably enough to make general inferences, it is still not large
enough to draw detailed statistical conclusions.

\section{Analysis of SN X-ray Lightcurves}
\label{sec:analysis}

A survey of the fits to the data in figure \ref{fig:plotlxfit} reveals
the diversity in the X-ray expansion of SNe. About half the SNe decay
with $\beta < 1$, a necessary (but certainly not sufficient) condition
if they are expanding in a steady wind. Two of them (SN 1987A and SN
1996cr) show $\beta < 0$, displaying X-ray emission that appears to
increase over long time-scales, as has been well documented in the
past \citep{cd95, mccray1993, park06, mccray2007, baueretal2008,
  ddb10}. Another two (SN 1979c and SN 2006jd) appear to be relatively
constant, although we caution that for SN 2006jd we have only two data
points, and are waiting for more data to be published. A few SNe have
$\beta > 1$. In the framework of the theory outlined above, these
could not arise from SN expansion into a steady wind. This group
consists of 4 Type IIn SNe, which tend to show some of the steepest
decreases in X-ray emission. Others in this category include SN 2003bg
and 2008ax, which do not have enough time-sampling over even a decade
to effective judge their luminosity decrease. The steepest light curve
belongs to SN 1970G, which drops as $t^{-3.08}$. Interestingly, many
of these SNe which have steep drop-off in light curves appear to be
those from whom X-ray emission was found only late in their lifetimes,
after about 6 years. In this context, we point out that if the X-ray
light curve of SN 1993J after 6 years is considered, the fall-off
would be much steeper than the gentler overall slope, as has been
shown in \citet{chandraetal09}.

We discuss some of our results in more details, in the context of
either individual SNe or groups of SNe.

\subsection{Type IIn Supernovae:} In 1990, a new class of SNe was introduced 
by \citet{Schlegel1990}. The 'n' denotes narrow lines, which
distinguish Type IIn from other SNe. Many of these SNe (but not all)
have a high radio and X-ray luminosity. In fact, as is evident from
Figure \ref{fig:plotlxtype}, as a class they have the highest X-ray
luminosity of all SN types.

As shown in Figure \ref{fig:plotlxfit}, 4 of the Type IIns (SN 1986J,
SN 1988Z, SN 1998S and SN 1995N) have a very steep luminosity
decline. {\em SN 1986J} X-ray flux decreases as $t^{-1.7}$. We note
here that for 1986J, we have adopted the flux values computed by
\citet{houck05} rather than those by \citet{Temple2005}. We recomputed
the values for the Chandra observations and find much closer agreement
with the former than the latter.

The steep value of the X-ray decline would suggest that the SN is
expanding in a medium with a density that declines faster than
$r^{-2}$. SN 1986J is also well studied in the radio range. VLBI
observations \citep{bietenholzetal10} suggest that the shock radius is
expanding as $R_{sh} \propto t^{0.69}$. The luminosity decline and
radial expansion then provide two equations which can be used to
determine the slope of the ejecta $n$ and the slope of the CSM density
profile $s$. It is easy to show however, that there is no finite
solution for $n$ and $s$ using equations \ref{eq:radexp} and
\ref{eq:lumband}. In other words, there is no compatible solution for
the case of an adiabatic shock in SN 1986J.

If we consider a radiative reverse shock, then it is possible to get a
finite solution which gives $s=2.72$. This appears to be consistent
with the luminosity decline, which indicates a very rapid
slope. However, it gives a value of $n=3.61$, which is not allowed
within the Chevalier solution \citep{chevalier1982a}.  Such a low
value of $n$ is inconsistent with most computed models of SN outer
ejecta density profiles \citep{cs89, matzner99}. That does not
completely exclude this solution. The emission could arise from the
reverse shocked material colliding with the flatter part of the ejecta
density profile (i.e. the self-similar solution is no longer
valid). It is also possible to envision scenarios such as the shock
becoming suddenly radiative, implying a transition stage, or expanding
into a medium that drops steeply in density.

These results seem to point to a complicated emission structure. An
alternative solution, as suggested by \citet{chugai93}, is that there
are two components to the emission. The basic idea is that of a SN
expanding in a clumpy wind, with the clumps having a low volume
filling factor. The expansion of the SN in the interclump wind medium
is mainly responsible for the radio emission and the radial expansion
measured with VLBI, whereas the X-ray emission arises from radiative
shocks within the clumps. However, this still implies that the density
of the clumps, or the number of clumps, must decrease faster than
$r^{-2}$ in order to get the appropriate decline.

Even if the SN is expanding into a clumpy wind, it is unlikely that it
would have been doing so since it exploded. If we assume the current
luminosity decline started right from the SN explosion, and
extrapolate the soft X-ray luminosity to 10 days after the explosion,
we get that it would be about 1.5 $\times 10^{44}$ ergs s$^{-1}$. This
is an extremely large luminosity, especially just in the soft X-ray
band, given that we would expect the hard X-ray luminosity to be much
larger at this stage (indeed, at the last Chandra observation, the
soft and hard luminosities were comparable). Note that if the SN
really was emitting in X-rays with this luminosity, it would have lost
all its kinetic energy in a year at that rate. Even given the rate of
luminosity decline it would have lost all its energy in a few
years. It is clear that such a high luminosity would be unsustainable.

The first X-ray detection of SN 1986J was made 8.43 years after
explosion, when the luminosity was almost 10$^{40}$ ergs s$^{-1}$. If
we assume that the SN maintained this constant luminosity since
explosion, it would mean that in the first 8.43 years it would have
lost a total of 2.5 $\times 10^{48}$ ergs just in the soft X-ray
band. The amount lost in hard X-rays would most likely be greater
since the temperature would be higher. The flux at infra-red
wavelengths can be up to a factor of 100 larger than that in X-rays,
especially at higher temperatures \citep{dweketal08}, thus leading to
an improbably high broadband flux. However it is possible that for SNe
expanding in a very dense medium, some of the dust may be destroyed,
and the IR flux may be only a few times larger than X-rays. The
broad-band flux would still be high, but not impossibly so.

A plausible inference is that the average luminosity of SN 1986J over
the first 8.43 years was lower than at 8.43 years, and perhaps orders
of magnitude lower. At some point before 8 years the X-ray luminosity
increased. In terms of the clumpy wind model, this could mean that
there were fewer to no clumps in the first few years, and that the
number of clumps increased after a few years. An alternative is that
the SN shock was evolving in a much lower density region earlier, and
started interacting with a higher density region at a later time. This
would imply a sudden deceleration in the shock velocity when it began
expanding into the high density region, which is not seen in the radio
VLBI data. We will explore the X-ray emission from SN 1986J in detail
in a future paper.

{\em SN 1988Z} has an X-ray luminosity that decreases even faster than
that of SN 1986J. Following the arguments above, a plausible model
appears to be that of a SN shock wave interacting with a clumpy wind,
as suggested by \citet{chugai94}.  Not surprisingly, extrapolating
back to 10 days, we get an X-ray luminosity of 6.5 $\times 10^{46}$
ergs s$^{-1}$. At this luminosity, the SN would lose all its energy in
one day, so its clear that the soft X-ray lightcurve could not have
had the recorded decline right from the time of explosion. Following
the line of reasoning above for SN 1986J, it appears most likely that
SN 1988Z also had a lower X-ray luminosity in the early years, which
increased at a later time. This then further suggests a lower density
medium, or lack of clumps in the ambient medium, during the first few
years.

{\em SN 1995N} and {\em SN 1998S} also have luminosity declining
faster than t$^{-1}$, although not as fast as 1986J or
1988Z. Nonetheless, it does suggest a medium that decreases faster
than r$^{-2}$. Extrapolating SN 1995N luminosity back to 10 days gives
a luminosity of $\sim 2.5 \times 10^{43} {\rm ergs}\; {\rm s}^{-1}$, still
larger than the observed X-ray luminosity of any known X-ray SN, and
unlikely (although not impossible), following the arguments given
above for SN 1986J. SN 1998S extrapolated back gives a luminosity of
$\sim 5.8 \times 10^{41} {\rm ergs} s^{-1}$, approaching the upper
limit of known X-ray luminosities. In the case of 1998S the decline is
not so steep, and it is possible that the luminosity could have
declined with this slope since the time of explosion.

The above slopes suggest that the density into which the SN is
expanding, whether it be due to clumps or a wind, falls faster than
r$^{-2}$. While most Type IIns show a very high luminosity, not all of
them decline so rapidly. SN 2005kd appears to rise and fall, over a
time-period of only a few 100 days. SN 2006jd appears to be more or
less constant over 2 data points separated by a few 100 days. Neither
of these would be consistent with an r$^{-2}$ density in the ambient
medium. SN 1996cr, which has been studied in detail
\citep{baueretal2008, ddb10, dbd11}, was increasing in X-ray
luminosity for several years, before turning over, and is thought to
have expanded in a low density region surrounded by a dense shell. 

It is clear that although there is great diversity, none of the Type
IIn appear to be evolving into a steady wind with constant parameters.

\subsection{SN 1993J:} This SN has more X-ray data than any other except 
SN 1987A.  The X-ray and radio emission has been studied in detail by
several authors, but no consensus reached on what kind of medium the
SN is expanding in. One of the earliest papers which studied the X-ray
light-curve \citep{sn95} suggested that the density immediately around
the SN was decreasing as r$^{-1.7}$.  A similar result, of a density
slope flatter than r$^{-2}$, was found by \citet{flc96}.  Following up
on a seminal paper that interpreted the radio emission of 1993J in the
context of a steady wind \citep{fb98}, \citet{fb05} asserted that both
the X-ray and radio lightcurves could be fit by a self-similar
expansion for SN 1993J, with a power-law ejecta expanding into an
r$^{-2}$ medium. They criticized the hydrodynamic modelling of
\citet{sn95} for using an ejecta profile that was hydrodynamically
unstable. However, \citet{nymarketal09}, citing the same hydrodynamic
models, suggested that the X-ray emission was coming partly from an
adiabatic shock expanding into CNO-enriched ejecta (which was not
described simply by a power-law), and partly from a radiative shock,
with no contribution from the circumstellar shock. The calculations of
\citet{chandraetal09} were consistent with the fact that the emission
was dominated by a reverse shock component, but did not need two
emitting components, although they did not study the spectra. The
radio and VLBI data has always been contentious, and the two different
views are summarized in \citet{barteletal07} and
\citet{martividaletal11}.

The soft X-ray light-curve for the first 700 days is well fit by a
slope $\beta = 0.34$. This could fit an $s=2$ solution if $n=5$, but
such an ejecta profile is inconsistent with explosion models
\citep{sn95}. Otherwise it cannot be fitted with a steady wind
expansion. As \citet{flc96} have shown, for $s \sim 1.7$ the fit is
good with a radiative shock solution. However, the behaviour after 700
days is much more complex and does not appear to fit a constant value
of $\beta$, thus suggesting that the slope is changing at this time. A
changing density slope was used in \citet{chandraetal09} to fit the
hard X-ray light curve. Overall, SN 1993J does not appear to fit the
expansion pattern of a shock expanding in a steady wind, especially
since the slope of the X-ray decay steepens at later times.

\subsection{SN 1979C:} This SN appears to be expanding with a  more or 
less constant luminosity for several years.  \citet{patnaudeetal11}
interpret this steady luminosity as evidence for emission from a
central black-hole component. Although without a detailed inspection
of the spectra, we cannot validate or dispute their assertion, we do
wish to mention here that having a more or less constant lightcurve is
not inconsistent with the circumstellar interaction model. In section
\ref{sec:xraysne} we have shown that for adiabatic shocks for example,
n=9, s=1.62 gives a constant luminosity in the soft X-ray band, while
for radiative shocks, $s=1.5$ gives a constant flux. Therefore having
a constant flux over an extended period in SN 1979C (and perhaps SN
2006jd) does not contradict the circumstellar interaction model,
although it does indicate that the evolution is not into a steady
wind.

\subsection{Other SNe:} SN 2006jc is an unusual SN that experienced a 
mass ejection about two years before becoming a SN
\citep{Pastorello2007, foley2007}. The X-ray emission
\citep{immleretal08} is consistent with the SN ejecta rising as it
impacts the dense shell resulting from this mass-loss, followed by a
decline as the shock crosses the dense region \citep[see for
  instance][]{Dwarkadas2005}. If it was interacting with an r$^{-2}$
wind, it was only for an extremely short period, but for the most part
the profile deviates significantly from a steady wind.

SN 1987A has been very well studied in the literature. Its proximity,
and the availability of the {\it Hubble Space Telescope}, has allowed
us for once to optically observe the interaction. Analysis of the
optical, x-ray and radio data have shown that the SN evolved into a
low density medium before impacting a higher density ionized HII
region \citep{cd95, park05, park06, parketal07,Dwarkadas2007d,
  Dwarkadas2007b} followed by a dense shell, part of a bipolar
circumstellar nebula. The evolution, and X-ray emission, is summarized
in reviews by \citet{mccray1993, mccray2003, mccray2007}.

SN 1978K also seems to be continually bright in X-rays. Its luminosity
could be considered constant on average, but appears to fluctuate by
factors of a few, making it difficult to get a good linear
fit. \citet{schlegel06} asserts that although there may not be too
much luminosity evolution, spectral evolution does exist. The current
X-ray light curve though appears inconsistent with a steady wind.

SN 1970G was discovered only several thousand days after
explosion. Although it currently has the steepest luminosity slope
$\beta =3.08$, it is not clear what the slope was over the first 20+
years. Currently the slope does not appear consistent with interaction
with an r$^{-2}$ medium.

\citet{schlegel01} analysed the X-ray emission from Type IIP SNe, and
suggested that they would not be strong X-ray sources. This is
consistent with the lightcurves in figure \ref{fig:plotlxtype}, which
shows that Type IIPs have a low X-ray luminosity. The fits to the
slopes of many of these, as well as other SNe not explicitly mentioned
above, are {\it theoretically} consistent with expansion within a
steady wind. The closest to that of a steady wind is that of SN
1994I. In most of these cases, there are either too few datapoints, or
the fits are not statistically good, so it is hard to judge if they
are evolving in a steady wind over an extended time period.  However,
\citet{cfn06} have studied the X-ray emission from Type IIP's assuming
they evolve in a steady wind, and their calculations tend to match the
observations.  It must be kept in mind that IIP's evolve from RSGs,
which do not have a very high wind velocity, and thus the wind will
not extend very far out from the star, perhaps a few parsecs at
most. Thus we would expect that even if the SN shocks were interacting
with winds, at some point early in their evolution (few hundred years
or less) they would run out of wind material to expand in.

\section{Discussion}
\label{sec:summary}

In this paper we have plotted the X-ray lightcurves of almost all
published X-ray SNe, in the narrow energy ranges in which they were
observed. Furthermore, we have fitted these lightcurves to a function
that goes as $t^{P(1)}$, and derived the value of $P(1)$, which we
have compared to theoretical expectations. In many of the cases, and
especially for Type IIn SNe, we find that the light-curve data in this
simple theory are not consistent with evolution into a steady wind
whose density decreases as r$^{-2}$. There are of course several
limitations to this theory. \citet{cf06} suggest that the x-ray
luminosities of type Ib/c SNe can only be explained by a non-thermal
mechanism, either X-ray synchrotron or inverse Compton. Our sample
does not include any GRB related 1b/c, and only a few non-GRB 1b/c SNe
that have published data. It is thus possible for scenarios to exist
where the light curve slopes, although not within the range outlined
in \S \ref{sec:xraysne}, may be consistent with a steady
wind. Although such exceptions are possible, it would be unusual to
assume that they would apply to all, or even most, SNe. Furthermore,
if the latter is indeed true, then it suggests that the many tens of
papers in the literature that have used this theory need to be
re-evaluated.

Our plotted lightcurves are consistent with those of
\citet{schlegel06} in those cases where overlap is possible. They are
inconsistent with those of \citet{immler07}, who plotted light curves
of all SNe known till then with just the first and last data point,
while asserting that ``all other X-ray data points are along the
extrapolated lines''.  As is evident from figure \ref{fig:plotlxall},
this does not seem to represent the available data. Using that
assertion, \citet{ki07} went a step further, calculating the density
profile of the medium into which the SN is evolving, and showing that
the results are consistent with a density decline corresponding to a
steady wind. Inherent in their calculation is that the SN luminosity
is decreasing as t$^{-1}$, which they take to mean that the ambient
density is going as $r^{-2}$, although as emphasized in \S
\ref{sec:xraysne} this is not necessarily true. Therefore, it is not
surprising that their final result matches that initial assumption.
Given the arguments above, and the light curves plotted in Figure
\ref{fig:plotlxall} and fitted in Figure \ref{fig:plotlxfit}, we do
not agree with their results.

Often in the literature, the X-ray lightcurve, or even a single X-ray
datapoint, has been measured, and used to compute a mass-loss rate,
assuming that the light curve decreases as $t^{-1}$
\citep{pooleyetal02, immleretal02, Immler2003, ik05, immler07, ki07,
  sp08, milleretal10b}. As shown above, and most clearly in
\citet{flc96}, the $t^{-1}$ dependence for a steady wind is only valid
over the total X-ray range, not the narrow bands in which SNe are
generally observed. Even if it were valid in the narrow range, none of
the published lightcurves are actually decreasing in luminosity as
t$^{-1}$, and very few are even close. We caution therefore that the
resulting mass-loss rates obtained in this manner from the X-ray
lightcurves may be incorrect, and not reflective of the true nature of
the CSM. If, as the lightcurves appear to indicate, the SNe are not
evolving in an r$^{-2}$ medium, then the calculated mass-loss rates
would need to be re-calibrated. Unfortunately, if the wind is not
steady, then the generally used expressions to measure the mass-loss
rates are no longer valid, and there is no simple way to compute the
mass-loss rate. Most techniques usually yield the combination
$\dot{M}/v_w$ - disentangling the magnitude, and time evolution, of
either or both parameters is a difficult task. If the deviation from a
steady wind is not large, a mass-loss rate obtained by assuming a
steady wind may be a good approximation early on, but the difference
increases gradually with increasing radius from the center of the
explosion.

An r$^{-2}$ density decline is often assumed in calculations of
mass-loss rates from optical observations \citep{smithetal08, scsff10,
  kieweetal10}. \citet{kieweetal10} have used this in fact to compute
the mass-loss rates for 15 type IIn SNe, and suggest that they are
consistent with LBV progenitors. It has not been shown that such a
density decline is warranted. \citet{salamanca03} pointed out many
years ago that the medium around Type IIn SNe does not necessarily
decrease as r$^{-2}$; this assertion was repeated for specific
individual cases by \citet{dwarkadas11}. If the nature of the density
distribution plays such an important part in determining the mass-loss
rate and thereby categorizing the SN progenitor, it is important that
the assumption of a steady wind density decreasing as r$^{-2}$ not be
made without accompanying proof.

There exist suggestions from other sources that the circumstellar
medium in the immediate vicinity of massive stars does not have an
r$^{-2}$ density profile, or that such a profile does not extend far
out. In their analysis of GRB afterglows, \citet{schulzeetal11} find
that only a quarter of them appear to be evolving in a freely
expanding wind, with the rest appearing to expand in a constant
density medium. Furthermore, in all cases except two, they place
limits on the freely expanding wind region of less than 1 pc, and $<$
0.1pc for about half the observed GRBs. Radio observations of SNe
appear to imply density modulations in the circumstellar medium
\citep{ryderetal04, soderbergetal06, ws11}, although its not clear how
large a deviation from a standard r$^{-2}$ profile is indicated.

X-ray emission from young SNe, which is mainly thermal, has the
simplest and most direct dependence on the density structure of the
surrounding medium, and can be used therefore to most easily infer
this density structure. In order to improve this analysis and make
more detailed calculations from SN X-ray lightcurves, it is imperative
that we have detailed lightcurves, with much better time sampling, and
over as wide an energy range as possible. Given the availability of
{\it Chandra}, {\it XMM-Newton} , {\it Suzaku}, and {\it Swift}, and
that they should continue operating for many more years, it is urgent
that we observe as many SNe as possible, and for as long as possible,
in X-rays.

\section*{Acknowledgments}
VVD would like to profusely thank John Houck and Dan Dewey (MKI) for
help with learning to use the ISIS software, and Dan for sharing many
of his scripts. A lot of the early observations listed herein are due
to Eric Schlegel, and many of the later ones to Dave Pooley and Stefan
Immler, to all of whom we are indebted.  We thank the anonymous
referee for a thorough reading of the manuscript and many insightful
comments. We are grateful to R. Chevalier for comments on an earlier
version of the paper. VVD's research is supported by grants TM9-0004X
(SN 1987A) and GO1-12095A (SN 1993J), both provided by the National
Aeronautics and Space Administration through Chandra Awards issued by
the {\it Chandra} X-ray Observatory Center, which is operated by the
Smithsonian Astrophysical Observatory for and on behalf of the
National Aeronautics Space Administration under contract
NAS8-03060. JG was supported by the UChicago NSF REU program in summer
2010.

\bibliographystyle{mn2e}
\bibliography{paper}


\bsp

\label{lastpage}

\end{document}